# Fungal Memory and Minimal Cognition

Kristina Šekrst

This paper argues that fungal mycelial networks exhibit minimal cognition through memory-integrated adaptive regulation. Drawing on cybernetic and enactivist frameworks, I develop a non-representational account of memory as the organism's capacity to modulate behavior based on temporally extended environmental coupling. I propose four operational criteria for minimal cognition: feedback-guided regulation of behavior, maintenance of internal viability conditions, structural modulation based on past environmental interactions, and plasticity across time scales that supports anticipatory adaptivity. Empirical evidence demonstrates that fungi meet all four criteria through distributed memory mechanisms: fungal networks exhibit directional regrowth toward previously encountered resources even after spatial displacement; stress priming persists across multiple cell divisions; Spitzenkörper-mediated directional persistence in constrained environments; and transgenerational memory through spore imprinting. These findings challenge representationalist assumptions in cognitive science by showing that memory and cognition can emerge from morphodynamic, biochemical, and electrophysiological processes without necessary neural substrates or symbolic representations. Fungal cognition demonstrates that the organizational principles underlying cognition – feedback-driven adaptation, norm-preservation, and historical coupling – can be realized in radically different material substrates, expanding our understanding of what counts as a cognitive system.

## 1. Introduction[1]

What counts as cognition? Traditionally, cognition has been closely associated with neural activity, implying that genuine cognition requires complex nervous systems capable of centralized information processing and behavioral control. However, recent advances in plant neurobiology[2] (Calvo, 2016; Calvo & Friston, 2017), studies of slime molds[3] (cf. Boussard et al.,

[1] To appear in *Topics in Cognitive Science*.
[2] I use the term "plant neurobiology" here descriptively, following Calvo (2016) and related literature, without endorsing any particular theoretical commitments about the nature of plant signaling systems. The term remains contested within plant science and philosophy of biology.
[3] Slime molds were once classified as fungi and are still often discussed alongside them, but they are now placed among the protists and are not true fungi. They recur in the minimal-cognition literature because, like fungal mycelia, they solve spatial problems through decentralized growth.

2021), and minimal cognition[4] frameworks (Calvo & Keijzer, 2011; Lyon, 2006) have challenged this neurocentric assumption. These perspectives argue that cognition may emerge in diverse biological architectures through adaptive, self-regulating organization, with memory constituting a fundamental dimension. In such frameworks, memory refers to persistent, functionally significant changes in physiological or morphological state that result from prior environmental interactions and influence subsequent behavior (Casadesús & Low, 2006) – operating through structural modifications rather than symbolic storage.

Fungal networks offer an especially compelling test case for this expanded conception of cognition. A mycelium is the vegetative body of a fungus, composed of branching filaments called hyphae that extend through soil, wood, or other substrates and transport water and nutrients across the colony. It grows and forages without any central organ of control, which is what makes it a useful case for theories of decentralized cognition. Fungal mycelia actively seek nutrient sources, adjust growth trajectories based on environmental conditions, and retain information[5] about past resource locations through directional regrowth and state-dependent morphological plasticity, and this kind of adaptive regulation emerges entirely without centralized control or neural tissue. The goal of this paper is to demonstrate that this type of memory-integrated adaptive self-regulation qualifies as minimal cognition exhibited by fungal networks.

Minimal cognition is operationalized here through four operational criteria synthesized from cybernetic and enactivist frameworks (developed in Section 2.4). (1) Feedback-guided regulation involves using sensory or internal signals to adjust behavior based on the outcomes of previous actions. (2) Maintenance of internal viability conditions describes the system's active regulation of essential variables (cf. Ashby, 1941, 1960), such as nutrient flow or structural integrity, to remain within the bounds required for continued existence. By “active” regulation, I refer to the system's endogenous modulation of internal states to maintain viability, not passive mechanical response, implying autonomous, self-organized regulation intrinsic to living systems. (3) Historical coupling is the structural or behavioral modulation of current behavior by past environmental interactions. (4) Plasticity across time scales supports anticipatory, future-oriented adaptivity. Together, these constitute minimal cognition: adaptive regulation that maintains internal viability constraints through feedback loops while systematically incorporating historical interactions into current behavior (cf. Barandiaran & Moreno, 2006; van Duijn, Keijzer & Franken, 2006), without requiring neural substrates or representational encoding. I use 'intelligence' and 'cognition' interchangeably throughout this paper, following the cybernetic tradition (Ashby, 1960) where both terms denote adaptive regulatory capacity – the operational focus is on the explicitly defined criteria for minimal cognition rather than intelligence as a distinct phenomenon.

[4] I retain the term “minimal cognition” despite Lyon's (2020) objection that it suggests a diluted form of “proper” cognition. Such use of “minimal” is methodological and not evaluative: it identifies systems meeting basic operational criteria for cognition (feedback-guided adaptivity, norm-preservation, historical coupling) without requiring neural structures: like “minimal cell” in synthetic biology, “minimal” denotes sufficient conditions, not deficient ones.
[5] I use "information" in a deflationary sense: talk of fungi "encoding" or "carrying" information is shorthand for persistent structural change that biases later behavior, with no commitment to stored representational content. The explanatory work here is done by historical coupling and viability regulation, not by information as a theoretical primitive. The term is nonetheless native to the cybernetic tradition the paper draws on, where Wiener (1948) treats it operationally, in terms of control and the reduction of uncertainty rather than representational content, and in that sense, it is consistent with the structural account developed here.

My claim is not that fungi possess cognition in the same sense as humans or other complex animals, but that fungal memory operates within a gradational framework in which even simple systems exhibit rudimentary forms of adaptivity and historical coupling (cf. Lyon, 2006; Barandiaran & Moreno, 2006). This narrower scope – anchored in regulatory and control principles, avoiding anthropomorphic projection – allows us to recognize meaningful cognitive features in fungal systems while providing valuable models for decentralized, self-organizing architectures.

The paper proceeds as follows: Section 2 establishes the theoretical framework by synthesizing cybernetic and enactivist approaches, developing a non-representational account of memory, and articulating four key criteria that distinguish cognitive from non-cognitive systems. Section 3 examines empirical evidence for fungal memory mechanisms – including directional regrowth, stress priming, and spore imprinting – illustrating how these mechanisms fulfill the given criteria. Section 4 returns to the theoretical philosophical framework discussed in Section 2 and considers the broader philosophical implications of fungal cognition for debates about representation, intentionality, and substrate-neutrality. Finally, I argue that fungal memory exemplifies genuine minimal cognition without necessarily requiring neural architecture and representations.

# 2. Theoretical framework

This section elaborates on the cybernetic and enactivist frameworks introduced above (2.1, 2.2), develops a non-representational account of memory (2.3), and establishes operational criteria for minimal cognition (2.4).

### 2.1 Cybernetics and adaptation

Cybernetics, as articulated by Norbert Wiener (1948), investigates the principles underlying control and communication in self-regulating systems. This approach emphasizes systemic relations among components rather than their physical identities: the system's organization, defined as the relational structure among parts, determines its functional dynamics (Ashby, 1960; Wiener, 1948). Cybernetic analysis thus adopts a substrate-neutral perspective, treating cognitive phenomena as properties of dynamic system organization and exemplifying this approach through analyses of systems that include humans, computers, animals, and even hybrid or artificial control systems.

W. Ross Ashby's (1941, 1960) contribution centered on the formal analysis of adaptation. He defined adaptive systems as those that maintain certain essential variables – such as temperature, pressure, pulse rate, etc. – within viable ranges despite external disturbances. An animal, for example, might regulate its body temperature within a narrow range despite fluctuations in the external environment, illustrating adaptation through feedback. The system reorganizes its internal structure to preserve these variables, continually comparing outputs against environmental inputs and adjusting when deviations occur.

When a system can react and also modify the conditions under which it reacts, adjusting its responsiveness parameters, Ashby called it “ultrastable” (Ashby, 1960, ch. 7). Ultrastability involves a second-order feedback loop operating on the system's internal regulatory mechanisms. This recursive architecture allows systems to survive in unpredictable environments by searching through configurations until stability is restored. Such systems need not be neural or even living: what matters is their dynamic organization to preserve internal coherence in the face of change (Ashby, 1941, p. 40).[6] That is, they must simply exhibit state-dependent modulation of internal structures in response to perturbations (Ashby, 1960, pp. 80–87). Phenomena like adaptivity, memory, and regulation can arise from purely functional principles instantiated in non-neural substrates.

Minimal cognition frameworks draw explicitly or implicitly on this cybernetic tradition (Barandiaran & Moreno, 2006; van Duijn, Keijzer & Franken, 2006). Yet Barandiaran and Moreno (2006) criticize early cybernetics (cf. Rosenblueth, Wiener & Bigelow, 1943) for reducing purposiveness to external behavior and feedback, lacking intrinsic teleology.[7] Following Jonas (1966/2001), they argue that genuine purpose depends on internal norms and not just on reactions to stimuli. Ashby's (1941, 1960) essential variables, however, already addressed this concern by grounding adaptivity in the internal regulation of viability constraints. Unlike external feedback mechanisms that respond only to environmental stimuli, essential variables are intrinsic parameters whose deviations beyond certain thresholds trigger endogenous reorganization, regardless of external conditions. This internalization of normative thresholds provides the "intrinsic teleology" Barandiaran and Moreno demand: the system regulates itself according to internally specified viability criteria, not merely in reaction to external goals. Ashby's ultrastability framework consequently shows that systems reorganize to preserve internal norms, aligning cybernetics with certain enactivist views of intrinsic purposiveness.

Van Duijn, Keijzer, and Franken (2006) developed a biologically grounded account of minimal cognition, positioning sensorimotor coordination as the critical factor distinguishing metabolic processes from cognitive phenomena. Minimal cognition cannot be reduced to isolated, centralized algorithms or neural mechanisms – it unfolds through a dynamic interplay between the organism and the environment. Consider, for example, the chemotaxis behavior (the directed movement up or down a chemical gradient) of *Escherichia coli* toward attractants or away from repellents: the bacterium's rod-like shape, fast perception pathway, and slower methylation feedback loops collectively enhance its capacity to navigate chemical gradients (van Duijn, Keijzer & Franken, 2006, p. 166). Morphology and temporal dynamics enable embodied, situated regulation of external conditions.

---

[6] Ashby explicitly emphasizes this functional unity between organism and environment, noting: "as the organism and its environment are to be treated as a single system, the dividing line between 'organism' and 'environment' becomes partly conceptual, and to that extent arbitrary. [...] but if we view the system functionally, ignoring purely anatomical facts as irrelevant, the division of the system into 'organism' and environment becomes vague" (Ashby, 1960, p. 40).

[7] I use "teleology" here in the sense of goal-directedness or purposiveness, acknowledging that some authors prefer "teleonomy" (Mayr, 1974) to distinguish naturally evolved functional organization from metaphysical purpose. My usage follows Barandiaran and Moreno's (2006) framework of intrinsic normativity.

By highlighting this type of organism-environment reciprocity, van Duijn, Keijzer, and Franken (2006) show that cognitive processes emerge through feedback loops that link and adjust both internal and external states. This allows incremental differentiation and expansion of sensorimotor abilities across evolutionary time, and the emphasis on sensorimotor coordination aligns closely with core cybernetic concepts – negative feedback,[8] maintenance of essential variables, and structural plasticity – all of which treat the organism and environment as facets of a single adaptive system. Minimal cognition emerges as reciprocal regulation rather than an exclusively internal, representation-based phenomenon, while the biogenic perspective (Lyon, 2006) analogously situates sensorimotor mechanisms as linking rudimentary metabolic functions with more advanced cognitive operations.

### 2.2 Enactivism and autonomy

Enactivist philosophy portrays cognition as an embodied, environmentally situated phenomenon (Di Paolo, Buhrmann & Barandiaran, 2017; Varela, Thompson & Rosch, 1991). Even though enactivism encompasses multiple schools of thought, including autopoietic enactivism (Varela, Thompson & Rosch, 1991), sensorimotor enactivism (O'Regan & Noë, 2001), and radical enactivism (Hutto & Myin, 2013), which differ in their treatments of representation and content, this approach primarily draws on the autopoietic tradition of Maturana and Varela's (1980) work and Di Paolo's (2005) development of autonomy and adaptivity, which align most closely with cybernetic concepts of norm-preservation and viability maintenance.

Enactivism situates cognition in reciprocal interplay between an organism's bodily organization and its surrounding environment. Here, autonomy refers not simply to self-organization present in various living and non-living systems but to the norm-generating capacity of a system to define and preserve its own identity across perturbation (Di Paolo, 2005). Norms are here defined functionally as viability constraints that trigger system reorganization when violated, compatible with multiple accounts of their origin. This capacity directly relates to the stability of the Ashbian (Ashby, 1941, 1960) essential variables and remains neutral on contested metaphysical debates (e.g., Millikan's (2021) teleosemantics vs. enactivist rejections of natural functions).

[8] Negative feedback regulates a system by counteracting deviations from a set point, reducing the difference between a target state and the actual state. Positive feedback amplifies deviations instead of correcting them.

Central to enactivism is historical coupling: the organism's capacity to integrate past environmental interactions into its ongoing regulatory dynamics, such that previous encounters modulate current and future behavior in ways that preserve viability. In other words, this is not memory as a standard kind of symbolic storage but as embodied modification: structural and physiological changes that tune the organism's responses based on its interaction history.

Di Paolo (2005) stresses that adaptive systems do not merely “happen” to stay alive but actively monitor and modulate internal states to mitigate threats and seize better conditions, offering a view that resonates with Ashby’s notion of ultrastability. Namely, both hold that systems must do more than compensate for external disturbances: they must reorganize internal parameters when first-order adjustments fail. As described above, essential variables must remain within viable thresholds to ensure the survival of the organism. If deviations exceed critical thresholds, the system must reorganize its internal structure: searching for new stable configurations to avert catastrophic failure.

Enactivism and ultrastability converge here: both treat cognition as self-regulatory, feedback-driven maintenance of crucial variables. Yet enactivism locates this regulatory dynamic specifically within living organisms, grounding it in biological autonomy as the minimal expression of cognition (Di Paolo, 2005). Autopoiesis, a concept developed by Maturana and Varela (1980), describes a system's continuous self-production and self-maintenance: autopoietic systems continually generate and conserve their organization, sustaining an identity over time. For enactivists, this self-maintaining organization provides the basis for biological autonomy, which in turn grounds minimal cognition (Di Paolo, 2005). Even the simplest life forms demonstrate basic cognitive processes by dynamically maintaining essential variables and reconfiguring themselves in response to environmental challenges.

Cybernetics and enactivism converge on norm-directed regulation but differ in scope. Cybernetics provides a formal, substrate-neutral framework applicable to any self-regulating system, living or artificial, while enactivism focuses specifically on biological autonomy and embodied interaction. How do these traditions relate? Ashby's (1941, 1960) essential variables specify the intrinsic viability constraints organisms must preserve. Autopoiesis (Maturana &

Varela, 1980) describes the self-producing organization that generates those constraints. Autonomy (Di Paolo, 2005) captures the system's capacity to regulate itself in accordance with self-generated norms. Ultrastability (Ashby, 1960) provides the mechanism for reorganization when regulation fails. Together, these concepts describe organisms as autonomous, autopoietic systems that use ultrastable regulation to maintain essential variables through norm-preserving environmental interaction.

### 2.3 Memory without representation

Classical cognitive science has typically conceived memory as the symbolic storage and retrieval of past states: internal representations that encode information about the world (Fodor, 1975; Lycan, 2023). Remembering involves accessing stored content that stands for or refers to previous experiences. However, this representational framework struggles to account for adaptive, history-sensitive behaviors in organisms without neural or symbolic systems. If memory requires internal representations, how can systems lacking the machinery for symbolic encoding nonetheless exhibit sophisticated regulation based on past interactions?

The anti-representationalist stance articulated here contends that cognition fundamentally arises from the organism's direct, history-dependent engagement with environmental contingencies. Drawing on enactivist philosophy (Di Paolo, Buhrmann & Barandiaran, 2017; Varela, Thompson & Rosch, 1991), adaptive behavior emerges as a process of historical coupling and self-maintenance. Organisms do not retrieve stored representations of the past; they adjust their embodied activities in ways shaped by previous interactions, all in the service of preserving their autonomy and vital norms (Thompson, 2007). Memory, in this framework, is therefore not the storage and retrieval of content but the organism's capacity to modulate current behavior through temporally extended environmental coupling.

A common objection to anti-representationalism is that internal states may still “stand in for” environmental features in a weak sense: that is, they function as quasi-representations even without a symbolic format (Lycan, 2023). While such deflationary views attempt to salvage representational language, they risk obscuring the core insight of enactivism: that cognition is constituted not by internal models but by ongoing regulatory interaction. When a system's structure changes due to environmental interactions and those changes bias future behavior, no component needs to correspond to a specific model of the world: structural modifications embody both past relevance and future viability. These modifications do not represent in the sense of encoding distal states; they enact meaning through viability-oriented change.

This anti-representationalist approach reframes intentionality itself as well. Traditionally, intentionality is described as “aboutness”: the idea that mental states encode or refer to something beyond themselves (cf. Searle, 1980). The anti-representationalist outlook defended here instead views intentionality as a system's norm-directed concern for its continued integrity (cf. Jonas, 1966/2001; Di Paolo, Buhrmann & Barandiaran, 2017). A living organism treats certain environmental factors as meaningful only insofar as they promote or threaten the system's essential variables (Ashby, 1941, 1960). When an organism reorients growth or activity toward previously beneficial conditions, it is not following a stored map of the environment but enacting a preference shaped by past success in maintaining viability. In this view, intentionality emerges as a relational property of organism-environment coupling, not as a property of internal states.

To avoid an overly permissive stance, I rely on strict operational criteria drawn from cybernetics and enactivism, requiring not merely structural responsiveness but norm-sensitive, history-dependent regulation oriented toward the maintenance of essential variables (Ashby, 1941, 1960), which distinguishes memory from simple material traces or mechanical hysteresis. A bent branch retains a trace of past forces, but this does not constitute memory because the trace plays no functional role in regulating the tree's future viability. Memory, by contrast, involves modifications that actively modulate behavior in service of norm-preservation. The key difference lies in functional integration: memory operates within a self-maintaining system that uses historical information to regulate essential variables.

This view addresses how cognition could be substrate-neutral and yet thoroughly embodied. Because representational mechanisms are not invoked, there is no requirement that adaptive intelligence must appear only where symbolic coding devices exist. Instead, the critical factors are the system's active maintenance of internal viability conditions and its capacity to flexibly reorganize its morphological or physiological configuration in light of historical interactions. Under these conditions, an organism can learn from experience without ever building or updating a repository of abstract mental content. Cognition is distributed across the organism's morphodynamic, biochemical, and electrophysiological processes. Ashby's (1960) insight that adaptive systems reorganize themselves to preserve essential variables thus follows the enactivist claim that organisms bring forth a lived world by continually regulating their own structural integrity in relation to environmental contingencies.

### 2.4 Operational criteria for minimal cognition

To avoid conceptual overreach, I propose the following operational criteria for identifying minimal cognition in biological systems, integrated from cybernetic (Ashby, 1960) and enactivist (Barandiaran & Moreno, 2006; van Duijn, Keijzer & Franken, 2006) frameworks:

1. feedback-guided regulation of behavior in response to environmental perturbation,
2. maintenance of internal viability conditions (essential variables),
3. structural or behavioral modulation based on past environmental interactions (i.e., historical coupling), and
4. plasticity across time scales that supports anticipatory or future-oriented adaptivity.

Why are these criteria not strictly sufficient? Cognition exists along a continuum of organizational complexity. I do not attempt to specify necessary and sufficient conditions for cognition in general, a task that remains contested in the philosophy of mind. Instead, these four criteria jointly identify the minimal threshold: systems meeting all four exhibit the basic organizational features distinguishing cognitive from merely reactive or homeostatic processes. Higher forms of cognition (in animals with nervous systems, for example) build upon these same foundational principles while adding capacities such as centralized processing, symbolic representation, predictive modeling, and, in some cases, conscious awareness. By focusing on minimal cognition, I establish a baseline against which more sophisticated variants can be compared, rather than offering a complete theory of cognition across all scales.

## 3. Fungal memory

Fungal memory operates through multiple, distributed mechanisms. At the cellular level, the Spitzenkörper – a vesicle-rich structure at hyphal tips that organizes cell growth and direction – integrates mechanical and chemical inputs into persistent morphological states (Held et al., 2019). At the molecular level, stress priming – the phenomenon in which a mild initial exposure to a stressor leaves an organism better able to withstand a later, more severe one – involves modifications in protein abundance, heat shock factors, and transcriptional profiles, suggesting that post-translational modifications and chromatin remodeling underlie lasting state transitions (Andrade-Linares et al., 2016).[9] Ion channel dynamics and bioelectric potentials may also contribute to the coordination of hyphal responses across spatial distances (Olsson & Hansson, 1995; Adamatzky & Gandia, 2021; Hunter, 2023), though fungal electrophysiology remains at an early stage, and the current evidence does not yet establish that electrical signaling plays a causal role in coordinating growth. These mechanisms collectively demonstrate that fungal memory is structurally and biochemically embedded, not ephemeral physiological noise. We will now examine how these mechanisms align with the operational criteria for minimal cognition.

### 3.1 Directional memory and spatial information

Mycelial networks are continuous, adaptive, and morphologically plastic, reorganizing themselves in response to both immediate perturbations and the lingering effects of past environmental conditions (cf. Boddy, 1999; Fricker, Boddy & Bebber, 2007). This plasticity reflects what Ashby (1960) described as ultrastability: systems that maintain viability by continuously adjusting their internal structure in response to external variability. Mycelial networks integrate environmental information and adjust growth or resource allocation based on past events – a form of spatial, distributed memory in a decentralized organism.

Fukasawa, Savoury, and Boddy (2020) demonstrated this using the wood-decaying fungus *Phanerochaete velutina*. They allowed a mycelium to grow from an initial wood block across the soil toward a new wood resource. After the fungus colonized the new wood, the researchers removed the original block and placed it in fresh soil, disconnecting it from the resource. When the fungus regrew from the relocated inoculum (the original colonized block), preferential growth occurred in the direction of the former resource – significantly more hyphae emerged on the side that had faced the bait. The mycelial network had stored a memory of the resource location, likely via lasting physiological changes oriented towards that direction. Notably, *P. velutina* would often entirely migrate into a sufficiently large new wood source, abandoning the original substrate; the threshold volume of the new resource (rather than the ratio of new vs. old) determined this relocation decision.

[9] The terms here name different levels of molecular change. Protein abundance is the quantity of a given protein in the cell. Heat shock factors are regulatory proteins that activate stress-response genes. Transcriptional profiles are the overall pattern of which genes are being transcribed at a given moment. Post-translational modifications are chemical changes made to a protein after it is synthesized, altering its activity or stability. Chromatin remodeling alters how tightly DNA is packaged and thereby changes which genes remain accessible for transcription. Each of these can persist after the triggering stimulus ends, providing a physical substrate for molecular memory.

However, this interpretation calls for two caveats. The directional regrowth could, in principle, arise from a topographic asymmetry rather than from memory, if the relocated inoculum carried more propagules (units capable of starting new growth, such as hyphal fragments) on the side that had faced the resource and so regrew preferentially in that direction. Fukasawa, Savoury, and Boddy (2020) themselves note that even neural memory is realized through physical changes in the structure of the network, and on the operational account adopted here, a persistent structural bias toward past resources is what memory consists in, not an alternative to it. The result has also not, to my knowledge, been independently replicated, so I treat it as suggestive rather than decisive, but the broader claim does not depend on it, since comparable structure-based spatial memory has been documented independently in *Physarum* (Kramar & Alim, 2021), where the geometry of the network itself encodes the history of nutrient distribution.

These behaviors constitute memory-integrated regulation similar to decision-making processes: the fungus assesses resource size and remembers spatial cues even after the resource is removed. The colony's network of hyphae acts as a distributed information system, where past foraging history influences future growth patterns. The ability to integrate past spatial information and modify growth accordingly exemplifies adaptive regulation: the fungus modulates future outputs based on feedback-mediated memory of prior inputs, thereby preserving essential variables, such as resource access, across time. This satisfies the enactivist condition of history-dependent regulation of viability, not merely in response to current stimuli but in anticipation of future interaction. Similar distributed memory mechanisms have been documented in *Physarum* slime molds, where tube diameter hierarchies encode spatial information about nutrient distribution (Kramar & Alim, 2021), suggesting that morphological memory may be a general principle in decentralized biological networks.

Recent work using live-cell imaging in microfluidic[10] environments shows that the Spitzenkörper serves as a cellular gyroscope, retaining directional memory in constrained geometries (Held et al., 2019). Microfluidics studies of *Neurospora crassa* show that the Spitzenkörper does more than coordinate local growth: it encodes directional persistence as memory at the level of intracellular organization. When a hyphal tip encounters an obstacle at an acute angle, the Spitzenkörper shifts laterally from its central position and maintains this asymmetry over extended distances, guiding the hypha along the contour of the obstruction. Microtubules, part of the cell's internal scaffolding (the cytoskeleton), reorganize in tandem, adopting skewed trajectories that reflect and preserve the prior growth vector, a pattern described as "cutting corners".

[10] Microfluidic devices contain channels at the scale of micrometers, used here to confine fungal growth within controlled geometries so that it can be imaged directly.

These structural biases actively conserve historical growth direction even after multiple deviations, and the cytoskeletal apparatus stores trajectory information as morphological inertia. When the Spitzenkörper is disrupted by frontal collision, however, this memory collapses and branching ensues, demonstrating the dependency of directional persistence on the integrity of the Spitzenkörper-microtubule system (Held et al., 2019). Fungal memory is therefore not only distributed but contingent, emergent from the mechanical continuity of self-organizing growth modules.

**3.2 Stress priming and molecular memory**

Another documented form of fungal memory is stress priming, in which a mild initial stress exposure primes the organism to withstand subsequent stressors more effectively. Stress priming (also known as "cross protection", "acquired stress response", "acclimation", "acquired stress resistance") has been identified in animals, plants, fungi, and bacteria (Andrade-Linares et al., 2016; Harish & Osherov, 2022; Harris, Amtmann & Ton, 2023; Hilker & Schmülling, 2019). Harish & Osherov (2022) describe this as an "adaptive response", which fits nicely within the Ashbian paradigm.

Multiple fungal species demonstrate this phenomenon. For example, Andrade-Linares et al. (2016) showed that soil filamentous fungi, species that grow as hyphal networks rather than as single cells, retain memory of heat stress. In their experiments, 19 soil fungi received a mild heat shock (35 °C) followed at various intervals by a higher heat stress (40 °C). Eight species exhibited a priming effect: prior heating significantly boosted their growth during the second stress compared to unprimed controls. Some fast-growing *Mucoromycotina* fungi showed higher growth after the second heat shock if primed, with a memory half-life of approximately 5–6 hours before the effect decayed. Similarly, *Penicillium chrysogenum* exposed to initial mild desiccation (loss of water from the tissue) later grew faster and showed higher enzyme activity under severe drought – priming benefits that persisted up to seven days (Guhr & Kircher, 2020). These and similar cases illustrate history-dependent change in fungal physiology: the memory of an earlier stress improves future performance, showing an adaptive trait in fluctuating environments.

Molecularly, stress priming involves lasting changes that outlive the initial stimulus. Memory may be stored in several forms (Andrade-Linares et al., 2016): altered gene expression, protein pools, metabolites (small molecules produced by metabolism), or epigenetic marks (chemical modifications that change gene activity without altering the DNA sequence). In *Saccharomyces cerevisiae*, a mild oxidative or heat shock induces cross-protection against other stresses, requiring new protein synthesis and persisting for several cell divisions – an epigenetic memory lasting up to five generations (Guan et al., 2012). The persistence, specificity, and adaptive orientation of these changes suggest a higher-order organization that exceeds mere reactive homeostasis. Following Ashby and enactivist principles, what distinguishes memory from simple acclimation is the integration of temporally extended information into regulatory behavior oriented toward systemic viability.

Alternative accounts interpret fungal stress priming as purely physiological acclimation rather than memory (cf. Wesener et al., 2021; Harish & Osherov, 2022). I acknowledge that physiological and cognitive framings may lie along a continuum rather than representing strict categories. However, the operational criteria established here – particularly historical coupling and anticipatory plasticity – provide clear functional distinctions between reactive homeostasis and memory-based regulation.

### 3.3 Transgenerational memory through spore imprinting

Fungal memory also spans life cycle stages. Conidia – asexual, non-motile spores – of filamentous fungi carry information about environmental conditions experienced during their formation, which subsequently affects their germination and growth. Kang et al. (2021) demonstrated this with *Aspergillus fumigatus*: they grew the fungus under nine different conditions (varying nutrients, temperature, oxidative stress, and metal availability), collected the spores, and tested them in various germination conditions. Despite being genetically identical, spores showed phenotypic differences depending on the environments from which they originated. *A. fumigatus* effectively imprints its spores with memory of parental stress, yielding offspring cells better suited to similar conditions (or more aggressive in a host). Recent work shows this primed state can be specific and long-lasting: conidia exposed to sublethal azole fungicide doses retain enhanced growth capacity for thirty days of storage, though the memory is transient across generations (Harish et al., 2022).

A study by Wang et al. (2021) uncovered a mechanism behind such spore priming. They found that mature conidia are not truly dormant but transcriptionally active before they detach from the parent fungus. In *Aspergillus nidulans*, *A. fumigatus*, and *Talaromyces* (*Penicillium*) *marneffei*, spores were shown to synthesize specific mRNAs in response to the conditions during their formation. These environment-specific transcripts (along with proteins and metabolites) are stored in the resting spore and can alter gene expression once germination begins. This bet-hedging strategy is a form of anticipatory regulation – a capacity to prepare for future environmental contingencies based on prior generative context (cf. Levy, Ziv & Siegal, 2012). The fungus hedges its bets by equipping spores with different “memories” of environmental cues so that at least some will match the conditions they eventually encounter. Such adaptive plasticity across a developmental transition highlights the breadth of fungal memory: not only do fungi remember within the somatic mycelium, but they also pass on memories to dispersal units (spores) to optimize survival in time and space.

### 3.4 Meeting the criteria for minimal cognition

From stress-hardening in yeast and molds to mycelial network decisions and spore priming, fungi demonstrate memory that meets the criteria established for minimal cognition. They retain history-sensitive physiological and morphological traces of past states, such as encountered temperatures or nutrient sources, which later modulate growth, metabolism, or development. Crucially, this memory is achieved without a central nervous system: it is an emergent property of molecular, cellular, and network-level processes (stored metabolites, epigenetic marks, persistent hyphal orientations, electrical signaling patterns). Fungal memory is distributed throughout the organism, consistent with the decentralized nature of a mycelium. Fungi use past interactions to regulate future behavior – a survival strategy demonstrating that cognitive processes can exist in unconventional substrates.

While critics might argue that memory in fungi is merely analogical rather than genuine, this distinction is rejected here. Following a longstanding operational tradition in cybernetics (Ashby, 1941), cognitive terms are defined by their functional role in regulatory systems: memory is not representational storage or recall, but rather the organism's capacity to modulate behavior through temporally extended environmental coupling. By this functional criterion, fungal memory *is* memory.

Fungal systems meet all four operational criteria for minimal cognition. First, feedback-guided regulation is evident: hyphal regrowth patterns shift in response to past nutrient locations (Fukasawa, Savoury & Boddy, 2020), reflecting dynamic feedback between stored information and current growth. This is negative feedback in the cybernetic sense: deviations from adequate resource access trigger compensatory growth responses toward resource-rich areas, maintaining essential variables (nutrient flow, metabolic activity) within viable thresholds (while, in the cybernetic sense, positive feedback would amplify such deviations rather than correcting them). Second, fungi maintain internal viability – turgor pressure, nutrient flow, osmotic balance – via morphological adaptation, such as migrating toward richer substrates or reorganizing network topology. Third, they modulate behavior based on past interactions: stress priming alters gene expression and growth profiles that persist beyond the initial trigger (Andrade-Linares et al., 2016; Guhr & Kircher, 2020), while directional regrowth after displacement reflects embodied spatial memory (Fukasawa, Savoury & Boddy, 2020). Fourth, fungal spores show plasticity across time scales: transcriptionally active conidia encode features of their formative environment (Wang et al., 2021), enabling anticipatory germination behavior and preparing offspring for future contingencies. Taken together, these examples demonstrate that fungal memory is an adaptive, history-sensitive regulation that preserves system viability, meeting all established criteria for minimal cognition.

# 4. Philosophical implications

Fungal memory provides empirical grounding for the anti-representationalist approach to cognition developed in Section 2.3. When a fungus reorients its mycelium toward previously nutritious locations, no component of the system corresponds to a "map" of the environment.

Rather, growth patterns embody past relevance and future viability through structural modifications: altered hyphal orientations, modified branching patterns, and shifted metabolic allocation. These changes bias future growth and resource transport without implying symbolic encoding. The fungus enacts a preference shaped by past success in maintaining viability, indicating how intentionality can be understood in normative and interactive terms rather than representational ones.

Fungal cognition exemplifies how intelligence can be substrate-neutral yet thoroughly embodied. Mycelial networks maintain morphological "traces" of past resource hotspots or stress exposures, and such traces bias future growth and resource transport in ways that exhibit memory without requiring symbolic code or centralized processing. An organism can learn from experience without building or updating a repository of abstract mental content. Cognition is distributed across morphodynamic, biochemical, and electrophysiological processes and not simply localized in representational structures.

This case is an excellent illustration of the relational nature of cognition. The mycelial system does not store or process standard symbolic information about nutrients or threats – it enacts a history-laden response to a changing environment. That is, instead of treating the mind (or memory module) as containing representations of external states, we can recognize that mind-like properties emerge from the interplay of self-maintaining dynamics (autopoiesis; Maturana & Varela, 1980) and adaptive modulations (Ashby, 1941, 1960) that refine the organism's responses. Fungal networks show that memory, agency, and intentionality can emerge from feedback-driven morphological plasticity without neurons or symbolic processing.

Several objections warrant consideration. First, one might argue that directional regrowth (Fukasawa, Savoury & Boddy, 2020) is merely refined tropism (directional growth toward or away from a stimulus) rather than genuine memory. However, tropisms respond to immediate environmental gradients, while directional persistence in fungi occurs even after displacement into new, neutral substrates. Second, critics may argue that stress priming is a transient physiological phenomenon. Yet the persistence of priming across multiple cell divisions and its specificity to prior exposures suggest regulatory plasticity with mnemonic features (Guan et al., 2012; Harish et al., 2022). Third, some might question whether fungal electrical activity represents anything beyond bioelectric noise. However, the regularity, stimulus-dependence, and multi-scale coordination of electrical spiking (Adamatzky & Gandia, 2021) point toward a role in signal propagation and network-level integration, analogous to primitive signaling in excitable tissues. These behaviors are best understood not as metaphors for cognition, but as minimal, embodied cases of it.

Recognizing fungal cognition as non-representational, dynamical intelligence broadens our understanding of cognition and reframes philosophical inquiry. Instead of searching for internal representations or mental content, we can examine how organisms preserve identity through structural regulation and analyze cognition wherever systems maintain norm-sensitive coupling with their environment. Fungal networks demonstrate that the organizational principles underlying cognition – feedback-driven adaptation, norm-preservation, historical coupling – can be realized across radically different material substrates.

## 5. Future directions

Fungal mycelial networks, though lacking neurons or centralized control, exhibit memory that meets operational criteria for minimal cognition. Drawing on cybernetics (Ashby, 1941, 1960) and enactivist philosophy (Di Paolo et al., 2017; Varela et al., 1991), I have shown that fungi display memory and adaptivity without necessary symbolic representation. Memory in these systems is distributed across morphological and biochemical structures – Spitzenkörper-mediated directional persistence, stress priming through epigenetic modification, and transgenerational spore imprinting – indicating that intelligence can emerge in diverse material substrates through norm-directed regulation.

This work challenges representationalist assumptions in cognitive science and the philosophy of mind. Fungal cognition expands the conceptual landscape of the mind, reinforcing a graded, pluralistic understanding of intelligence. Neural circuitry is not mandatory for sophisticated, adaptive behavior. By foregrounding regulatory dynamics, material plasticity, and historical coupling, I offer a substrate-neutral framework for understanding memory and intelligence: one applicable to life forms beyond the brain and architectures beyond the neuron. That fungi exhibit cognition in the minimal, morphodynamic sense does not diminish cognition but expands its horizon. Fungal networks invite us to rethink what it means to be a cognitive agent in a world built not from thoughts, but from flows, networks, and histories sedimented in form.